# Simultaneous summation and recognition effects for a dual-emitter light-induced neuromorphic device

Yongchao Yang, Bingcheng Zhu, Jialei Yuan, Guixia Zhu, Xumin Gao, Yuanhang Li, Zheng Shi, Zhiyu Zhang, Yuhuai Liu, and Yongjin Wang

**Abstract-We propose and fabricate a dual-emitter light-induced neuromorphic device composed of two light-induced devices with a common collector and base. Two InGaN multiple quantum well diodes (MQW-diodes) are used as the emitters to generate light, and one InGaN MQW-diode is used as the common collector to absorb the emitted light. When the presynaptic voltages are synchronously applied to the two emitters, the collector demonstrates an adding together of the excitatory post synaptic voltage (EPSV). The width and period of the two input signals constitute the code to generate spatial summation and recognition effects at the same time. Experimental results confirm that temporal summation caused by the repetitive-pulse facilitation could significantly strengthen the spatial summation effect due to the adding together behavior when the repetitive stimulations are applied to the two emitters in rapid succession. Particularly, the resonant summation effect occurs at the co-summation region when the two repetitive-pulse signals have a resonant period, which offers a more sophisticated spatiotemporal EPSV summation function for the dual-emitter neuromorphic device.**

## I. INTRODUCTION

The artificial synaptic device is a hot topic for brain-inspired neuromorphic systems [1-5], and synaptic electronics have gained considerable attention in recent years, including two-terminal memristors and three-terminal ionic/electronic hybrid devices. Based on phase change materials, nanoelectronic programmable synapses have been developed for brain-inspired computing [6]. Dynamic logic and learning have been presented using a carbon nanotube synapse [7]. An $Ag_2S$ inorganic synapse has been reported to emulate the synaptic functions of both short-term plasticity and long-term potentiation characteristics [8]. Flexible metal oxide/graphene oxide hybrid neuromorphic devices have demonstrated the realization of spatiotemporal correlated logics [9]. Proton-conducting graphene oxide-coupled neuron devices have been proposed for brain-inspired cognitive systems [10]. Compared with other synaptic devices, the light-induced synaptic device uses photons rather than electrons or protons to induce excitatory postsynaptic voltage (EPSV) behavior for artificial synapse applications [11].

Here, we propose and fabricate a dual-emitter light-induced neuromorphic device on an III-nitride-on-silicon platform. Figure 1(a) shows a schematic illustration of the proposed dual-emitter light-induced neuromorphic device, which has a common base (B). The collector (C) absorbs the pulse light generated by the emitter (E) to achieve a photon-electron conversion, leading to an EPSV for the mimicking of synaptic activity with different signal sources. The period, shape and width of pulses constitute the code to transfer information in the biological nervous system, which is characterized by the EPSV summation [12]. When the two emitters are synchronously biased, the EPSVs happen at the same time and are added together, leading to a spatial summation. The adding together of EPSVs generated at the same emitter forms a temporal summation if they occur in a rapid succession. The spatial summation can be significantly reinforced by the temporal summation, which is investigated for emulating the complicated memory effect during the learning process.

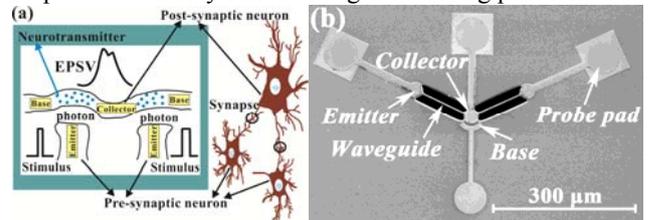

Fig. 1. (a) Schematic diagram of a dual-emitter neuromorphic device; (b) SEM image of fabricated dual-emitter neuromorphic device.

## II. EXPERIMENTAL RESULTS AND DISCUSSION

The proposed dual-emitter neuromorphic device is fabricated on a 2-inch III-nitride-on-silicon wafer. The 1500-μm-thick starting wafer is firstly thinned to approximately 200 μm by chemical mechanical polishing. The emitter, collector and probing pad are patterned by photolithography and formed by induced coupled plasma reactive ion etching (ICP-RIE) of III-nitride epitaxial films with $Cl_2$ and $BCl_3$ hybrid gases at the flow rates of 10 sccm and 25 sccm, respectively. The Ni/Au (20nm/180nm) metal stacks are used as p- and n-type contacts. Then, waveguide structures are defined and etched by ICP-RIE. After protecting the top device structure with thick photoresist, silicon removal is conducted by deep reactive ion etching with alternating steps of $SF_6$ etching and $C_4F_8/O_2$ passivation. Subsequently, III-nitride backside thinning is carried out by ICP-RIE to obtain ultrathin membrane-type device architecture. Figure 1(b) shows a scanning electron microscope (SEM) image of the fabricated dual-emitter neuromorphic device. The suspended device architecture can form a highly-confined waveguide structure for the in-plane light coupling between the emitter and the collector. Two InGaN multiple-quantum-well diodes (MQW-diodes) are used as the emitters to generate light and one InGaN MQW-diode serves as a common collector to absorb the in-plane guided light through suspended waveguides. The two emitters are connected to the collector via two 100-μm-long, 2.47-μm high, and 6-μm-wide suspended waveguides, and the electrodes are connected to the 70-μm-diameter probe pads for device characterization.

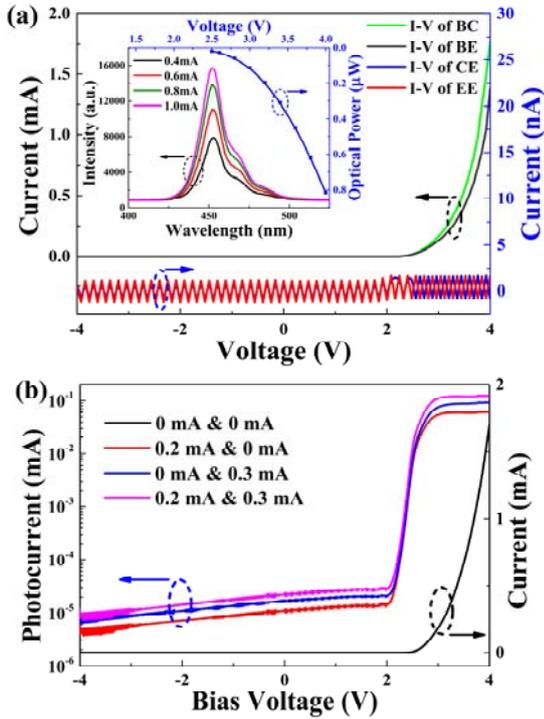

Fig. 2. (a) I-V curves of the four-terminal device, and the inset is electroluminescence (EL) spectra and optical power of the emitter; (b) Induced collector photocurrent versus forward current of the emitter.

Figure 2(a) shows the measured EL spectra of single light emitter at different injection currents. The light emission intensity depends on the injection current, and the dominant emission wavelength locates around 452 nm, as shown in the inset of Fig. 2(a). The optical output power from single light emitter is controlled by the injection current of the MQW-emitter. The MQW-collector absorbs the emitted light to generate the photocurrent. The base contact is probably 150 μm away from emitter contact, and there will be a large resistance in BE. As a result, the current-voltage (I-V) curve of BE shows a relatively small current. According to the I-V curve of EE, the electrical isolation is obtained, indirectly indicating that the signal is coupled optically. The I-V curve of BC exhibits a typical rectifying behavior without light absorption, which is used to normalize the induced photocurrent with light absorption. When the two emitters are separately driven by the injection currents of 0mA&0.2mA to emit light, Figure 2(b) shows an increased photocurrent with increasing bias voltage of the collector from -4 to 4 V, in which the collector exhibits two light detection modes. Because the two emitters are symmetrical, an increased photocurrent is observed when the two emitters are synchronously driven by the injection currents of 0.3mA&0mA. The light intensity is modulated by the injection current of the emitter, so the induced photocurrent is dependent on the injection current of the emitter. Thus, the generated photons from different emitters can be accumulated when both emitters are operating simultaneously. The induced photocurrent is added together when both emitters simultaneously operate with the injection currents of 0.3mA&0.2mA.

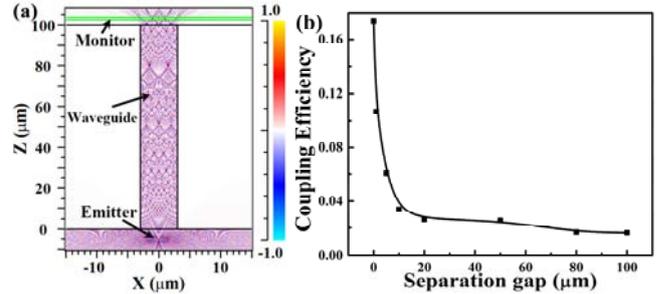

Fig. 3. (a) Light coupling without separation gap; (c) coupling efficiency as a function of separation gap.

Figure 3(a) shows the calculated light coupling between the emitter and the collector through suspended waveguide at a wavelength of 452 nm using beam propagation simulation, in which the refractive index of waveguide used is 2.45 and suspended waveguide is 100 μm long and 6 μm wide. Assuming that the suspended waveguide has a separation gap in the middle, Figure 3(b) shows the calculated coupling efficiency as a function of the separation gap. The coupling efficiency is significantly decreased with increasing the separation gap from 0 to 100 μm, indicating that the light being coupled in the waveguide does play a dominant role in the light coupling between the emitter and the collector because these optical components are fabricated on the same membrane [13].

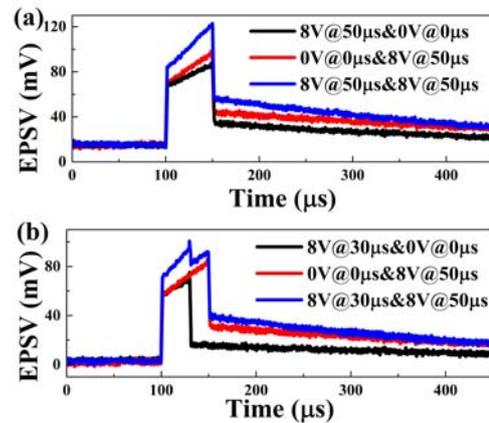

Fig. 4. Spatial EPSV summation: (a) same width; (b) different widths.

Figure 4(a) shows the spatial summation of the collector, in which two 8.0V@50μs pulse signals are separately applied to the two emitters simultaneously. Both the collector and the base operate at zero bias, and the two emitters are biased at 2.4 V. Compared to the EPSV generated by a single pulse signal from one emitter, the integrated EPSV amplitude increases, and the decay time extends, leading to an improved spatial summation effect. Moreover, the adding together of EPSVs is dependent on the pulse widths, as illustrated in Fig. 4(b). One 8.0V@50μs pulse signal and one 8.0V@30μs pulse signal are synchronously applied to the two emitters. The EPSV is integrated when two emitters are operating. The EPSV is then abruptly dropped and gradually increases when one pulse signal is terminated and only one emitter is driven by another pulse signal. Finally, the EPSV amplitude gradually returns towards its initial state due to the decay behavior when the later pulse signal is also ended. The EPSV summation can enhance the signal amplitudes and remain their differences, and the integrated EPSV can be decoded into simple signals that distinctly identify the light source, leading to a recognition function. The dual-emitter light-induced device can achieve signal recognition through a spatial EPSV summation process when the signal pulse widths are different.

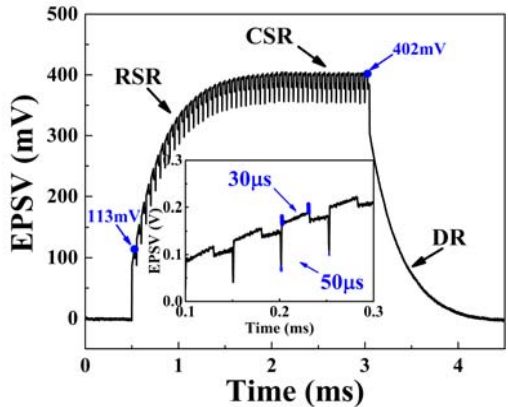

Fig. 5. Spatiotemporal summation under different pulse widths.

Repetitive-pulse facilitation (RPF) behavior occurs when the stimulation is continuously applied. Many EPSVs add together to produce a temporally integrated EPSV and finally reach a saturated value, which is determined by the initial amplitude of the pulse signal. However, it is still difficult to obtain a high temporal EPSV summation for the desired integration effect even if weak pulse signals are continuously applied [14]. The spatiotemporal EPSV can be achieved with the improved EPSV behavior through an adding together of spatial EPSV summation in a dual-emitter light-induced neuromorphic device. Two repetitive-pulse signals are separately applied to the two emitters simultaneously, in which the same pulse numbers are 50 and the same repetitive-pulse periods are 51 μs with the pulse widths of 50 and 21 μs, respectively. Figure 5 shows that the spatiotemporal EPSV amplitude increases and is eventually saturated as the pulse number increases. The spatiotemporal EPSV is improved due to the adding together of temporal and spatial summation. Compared to the first integrated EPSV amplitude of 113 mV, the saturated EPSV value is significantly increased to 402 mV. Signal recognition can simultaneously be obtained when repetitive stimulations with different pulse widths happen at the same period.

The spatiotemporal summation for a dual-emitter light-induced device is highly sensitive to the period of the repetitive-pulse signals. Compared to the spatial summation of two single signals from two emitters, the spatiotemporal summation behavior can be divided into a rapid summation region (RSR), a co-summation region (CSR), a temporal summation region (TSR), and a decay region (DR). Figure 6 shows the spatiotemporal summation behavior at different signal periods. One signal has a pulse width of 50 μs and a signal period of 51 μs, and the other has pulse width of 20 μs and a signal period of 21 μs. The pulse numbers of two repetitive signals are 50. Two RPF behavior patterns occur at RSR, and the integrated EPSV amplitude is quickly increased.

As the pulse number increases, the integrated EPSV amplitude enters into the CSR, in which the difference between the two signals leads to the signal recognition at a high EPSV level. When one repetitive signal is ended, the co-summation effect is stopped and the saturated EPSV amplitude is quickly dropped and enters into the TSR, in which only one RPF takes place. When the later signals are finished, the EPSV gradually returns to the initial state, resulting in a decay process. When the numbers of the two repetitive signals are increased, a periodic EPSV summation occurs, which relates to the resonant period. This sophisticated summation function occurs when the two repetitive pulse signals have a resonant period in the co-summation process. The two repetitive-pulse signals have the signal periods of 51 μs and 21 μs, respectively. Therefore, the two repetitive-pulse signals have a resonant period of 357 μs, leading to a distinct resonant summation effect (RSE). Taken together, it can be concluded that the RSE provides a more complicated summation function for a multiple-emitter device, resulting in the simultaneous summation and recognition of multiple signals at a high level.

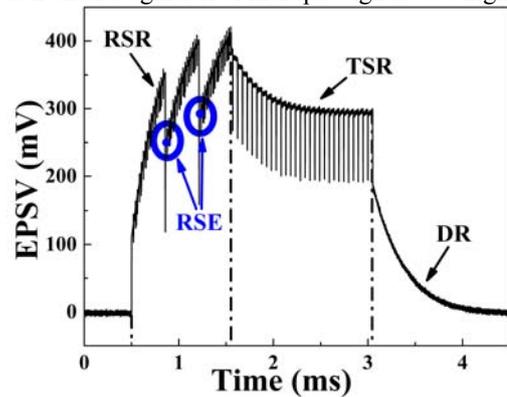

Fig. 6. Spatiotemporal summation behavior under different signal period.

### III. CONCLUSION

In conclusion, we propose and fabricate a dual-emitter light-induced neuromorphic device to investigate the adding together of temporal and spatial EPSV summation effect. The integrated EPSV behavior occurs when pulse signals are synchronously applied to the two emitters. The period and width of signals lead to improved summation and recognition behavior at the same time. Experimental results confirm that the summation effect could be significantly strengthened due to the adding together of EPSV summation when repetitive stimulations are applied to the two emitters. Particularly, the RSE occurs at the CSR when the two repetitive-pulse signals have a resonant period, which provides a more sophisticated spatiotemporal EPSV summation function.